\documentclass[groupedaddress,superscriptaddress,showpacs,notitlepage,11pt]{revtex4-1}

\usepackage{graphicx}
\usepackage{sidecap}
\usepackage{makecell}
\usepackage{array}
\usepackage{multirow}
\usepackage{amsmath}

\newcolumntype{x}[1]{>{\centering\hspace{0pt}}p{#1}}

\def\dj{d\kern-0.4em\char"16\kern-0.1em}
\def\Dj{\mbox{\raise0.3ex\hbox{-}\kern-0.4em D}}

\begin{document}

\title{Lattice dynamics of FeSb$_{2}$}

\author{N. Lazarevi\'c}
 \affiliation{Center for Solid State Physics and New Materials, Institute of Physics Belgrade, University of Belgrade, Pregrevica 118, 11080 Belgrade, Serbia}
\author{M. M. Radonji\'c}
 \affiliation{Scientific Computing Laboratory, Institute of Physics Belgrade, University of Belgrade, Pregrevica 118, 11080 Belgrade, Serbia}
\author{D. Tanaskovi\'c}
 \affiliation{Scientific Computing Laboratory, Institute of Physics Belgrade,
University of Belgrade, Pregrevica 118, 11080 Belgrade, Serbia}
\author{Rongwei Hu$^{\ast }$}
 \affiliation{Condensed Matter Physics and Materials Science Department, Brookhaven
National Laboratory, Upton, New York 11973-5000, USA}
\author{C. Petrovic}
 \affiliation{Condensed Matter Physics and Materials Science Department, Brookhaven
National Laboratory, Upton, New York 11973-5000, USA}
\author{Z. V. Popovi\'c}
 \affiliation{Center for Solid State Physics and New Materials, Institute of Physics Belgrade, University of Belgrade,
Pregrevica 118, 11080 Belgrade, Serbia}

\begin{abstract}
The lattice dynamics of FeSb$_2$ is investigated by the first-principles DFT calculations and Raman spectroscopy. All Raman and infrared active phonon modes are properly assigned. The calculated and measured phonon energies are in good agreement. We have observed strong mixing of the A$_g$ symmetry modes, with the intensity exchange in the temperature range between 210 K and 260 K. The A$_g$ modes repulsion increases by doping FeSb$_2$ with Co, with no signatures of the electron-phonon interaction for these modes.
\end{abstract}

\pacs{ 63.20.D-; 71.15.Mb; 71.28.+d; 78.30.Hv; }
\maketitle

\section{Introduction}

FeSb$_{2}$ is a strongly correlated narrow-gap
semiconductor which has recently attracted a lot of attention due to its
unusual thermoelectric properties.\cite{1,2,3,4,5,6} It was shown that FeSb$_{2}$ has colossal thermopower $S$ at 10
K (range from 1 mV/K to 45 mV/K\cite{1,3.1}) and the largest power factor $S^{2}\sigma $ ever reported.\cite{1,3.1,3.2,3.3}
The phonon contribution to $S$ remains controversial.\cite{6.1} Also, the
thermal conductivity $\kappa $ of FeSb$_{2}$ is relatively high and around 10 K
is dominated by the phonons.\cite{1} Consequently, full knowledge of
FeSb$_{2}$ lattice dynamics is necessary in order to understand the low temperature transport and thermodynamic properties of this material.

The infrared active phonon frequencies of FeSb$_2$ were obtained from the polarized far-infrared reflectivity spectra.\cite{10.1} From E$\|$b
polarized reflectivity measurements on (102) plane of FeSb$_{2}$ single crystal, Perucchi \emph{et al.} observed four modes at 106.4, 231, 257 and 271 cm$^{-1}$ at
10 K (factor group analysis predicts 3$B_{2u}$ modes for this
polarization configuration). For E$\bot$b polarization, both $B_{1u}$ and B$_{3u}$
symmetry modes can be observed from (102) plane. Three (of four)
modes at 121, 216 and 261.4 cm$^{-1}$ are observed for this
polarization. Raman scattering measurements on FeSb$_2$
were published in Refs. \onlinecite{7.a,7.b,7,7.1}. Lutz and
M\"{u}ller\cite{7.a} observed two Raman active modes at about 175
and 154 cm$^{-1}$ on hot-pressed samples, and assigned them as the A$_{g}$ symmetry modes.
In contrast, Racu \emph{et al.}\cite{7.b} observed three
Raman modes at about 150, 157 and 180 cm$^{-1}$ using polarized Raman scattering measurements on FeSb$_2$ single crystals and assigned them as the B$_{1g}$, A$_{g}$ and B$_{1g}$ symmetry modes, respectively. Finally, all six Raman
active modes of FeSb$_{2}$ predicted by the factor group analysis
(2A$_{g}$+2B$_{1g}$+B$_{2g}$+B$_{3g}$) were observed in Ref. \onlinecite{7}. Polarized Raman scattering spectra of the Fe$_{1-x}$M$_x$Sb$_2$ (M=Cr,Co) single crystals was studied in Ref.~\onlinecite{7.1}. The linewidths and energies of the Raman
modes were analyzed as a function of doping $x$ and temperature.
Strong electron-phonon interaction, observed for the B$_{1g}$
symmetry mode of pure FeSb$_2$, produces significant mode
asymmetry. The electron-phonon interaction is drastically reduced with increasing concentration of Co and Cr in
Fe$_{1-x}$(Co,Cr)$_x$Sb$_2$. The mixing of the A$_{g}$ symmetry phonon
modes has been observed both in pure and Cr-doped samples.\cite{7.1}

In this paper we report \emph{ab initio} study of the lattice dynamics of FeSb$_2$. The calculated phonon energies in the
$\Gamma$ point are in good agreement with experimental data. Phonon density of state show a gap at about 175 cm$^{-1}$, which divides a low frequency region where vibration modes are mostly Raman active from a high frequency region where only infrared active modes appear. The calculated phonon dispersions for two A$_g$ symmetry modes indicates strong mode mixing. This is indeed observed in our polarized Raman scattering spectra. The A$_{g}$ mode intensity exchange in the temperature range between 210 K and 260 K agrees well with theoretical calculations, excluding any additional temperature dependent electron phonon coupling for these modes. The mode repulsion increases with Co doping.

\begin{figure*}
\includegraphics[height=1.2\textwidth, angle=0]{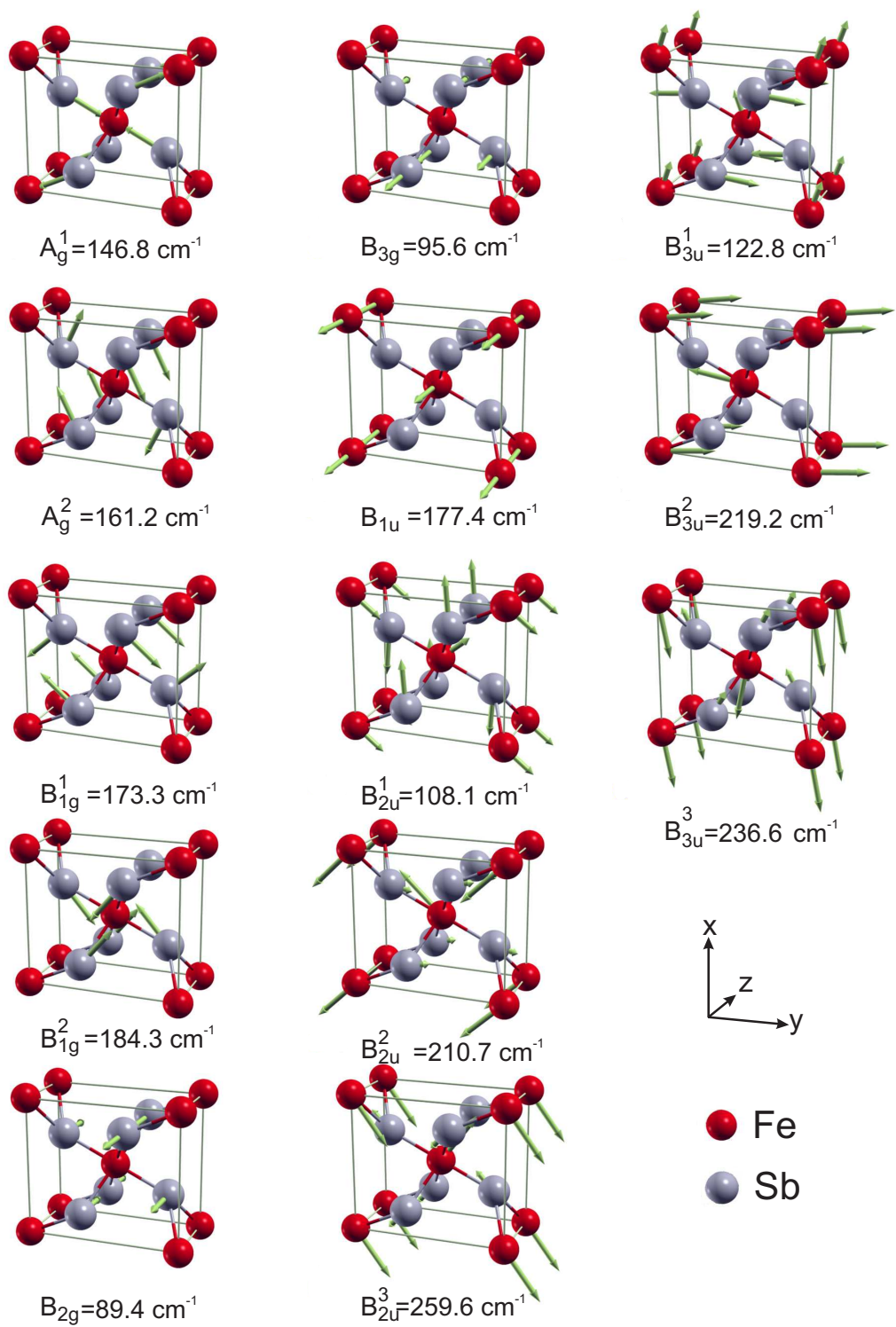}
\caption{(Color online) Atomic displacement patterns for the vibrational modes of FeSb$_2$. The length of the arrows are proportional to square root of the vibration amplitudes.} \label{fig1}
\end{figure*}

\section{Experiment}

Single crystals of FeSb$_{2}$ and Fe$_{0.75}$Co$_{0.25}$Sb$_{2}$
were grown by the self-flux method and characterized as described elsewhere.\cite{2} The Raman scattering measurements were performed using Jobin Yvon T64000 Raman system in micro-Raman configuration.
The 514.5 nm line of an Ar$^{+}$/Kr$^{+}$ mixed gas laser was used
as an excitation source. Focusing of the laser beam was realized
with a long distance microscope objective (magnification $50\times
$).  We have found that laser power level of 0.02 mW on the sample
is sufficient to obtain Raman signal and, except signal to noise
ratio, no changes of the spectra were observed as a consequence of
laser heating by further lowering laser power. The corresponding
excitation power density was less then 0.1 kW/cm$^{2}$. Low
temperature measurements were performed between 15 K and 300 K using
KONTI CryoVac continuous Helium flow cryostat with 0.5 mm thick
window. Raman scattering measurements of pure and Co doped FeSb$_2$
samples were performed using the
$(10\bar{1})$ oriented FeSb$_2$ samples.
Selection rules for parallel and crossed polarization from the
$(10\bar{1})$ plane of the orthorhombic crystal symmetry and the mode assignment have been presented in Ref.~\onlinecite{7}.

\section{Numerical method}

FeSb$_{2}$ crystallizes in the orthorhombic marcasite-type structure of the centrosymetric Pnnm (D$_{2h}^{12}$)
space group, with two formula units (Z=2) per unit cell.\cite{31,32} Basic structural unit is built up of Fe atoms surrounded
by deformed Sb octahedra. These structural units are corner sharing
in the (ab) plane and edge sharing along the $c$-axis. Two Fe
atoms are in (2a) Wyckoff positions at (0,0,0) and four Sb atoms are in (4g)
Wyckoff positions at (0,$u$,$v$) of the Pnnm space group. Our density functional theory (DFT) calculations are
performed within generalized gradient approximation (GGA) with PW91 exchange-correlation functional which is used to
calculate ultra-soft pseudopotentials,\cite{21.1} as implemented in the QUANTUM ESPRESSO package.\cite{21}  Iron (antimony) pseudopotential takes into account 3s$^2$ 3p$^6$ 4s$^2$ 3d$^6$
(4d$^{10}$ 5s$^2$ 5p$^3$) electron states for the valence electrons. The Brillouin zone was
sampled with an 8$\times$8$\times$8 Monkhorst-Pack \textbf{k}-space
mesh and with the Marzari-Vanderbilt cold smearing (0.005Ry).\cite{21.2} The obtained optimized structural parameters are $a$=5.859 \AA, $b$=6.583
\AA, $c$=3.812 \AA, $u$=0.1882, and $v$=0.3554, which are in
good agreement with the experiment. Our band structure calculations agree well with the previously reported. \cite{6.1,6.a,6.b}

\section{Results and discussion}

\begin{table*}[t]
\caption{Raman and infra-red active mode energies (in cm$^{-1}$) of FeSb$_2$
single crystal.}
\label{tab.1}
\begin{ruledtabular}
\setcellgapes{2pt}
\makegapedcells
\centering
\begin{tabular}{x{1.5cm}x{1.5cm}x{1.cm}x{1.7cm}x{1.9cm}|x{1.5cm}x{1.5cm}x{1.7cm}x{1.5cm}}
Symmetry   & Exp.\cite{7} &$\Omega(0)$ & Calculation & Activity & {Symmetry}  & Exp.\cite{10.1,10.2}  & Calculation & Activity\tabularnewline\hline
A$_g^1$    & 150.7         &  160.3     & 146.8       & $R$      &{B$_{1u}$}   & 195   & 177.4       & $IR$     \tabularnewline
A$_g^2$    & 153.6         &  164.4     & 161.2       & $R$      &{B$_{2u}^1$} & 106.4 & 108.1       & $IR$     \tabularnewline
B$_{1g}^1$ & 154.3         &  164.6     & 173.3       & $R$      &{B$_{2u}^2$} & 231.0 & 210.7       & $IR$     \tabularnewline
B$_{1g}^2$ & 173.9         &  190.4     & 184.3       & $R$      &{B$_{2u}^3$} & 257.0 & 259.6       & $IR$     \tabularnewline
B$_{2g}$   &  90.4         &            &  89.4       & $R$      &             & 271.0 &             & $IR$     \tabularnewline
B$_{3g}$   &  151.7        &            & 95.6        & $R$      &{B$_{3u}^1$} & 121.0 & 122.8       & $IR$     \tabularnewline
           &               &            &             &          &{B$_{3u}^2$} & 216.0 & 219.2       & $IR$
           \tabularnewline
           &               &            &             &          &{B$_{3u}^3$} & 261.4 & 236.6       & $IR$     \tabularnewline
           &               &            &             &          &{A$_{u}^1$}  &       &  84.9       & $Silent$\tabularnewline
           &               &            &             &          &{A$_{u}^2$}  &       & 195.2       & $Silent$\tabularnewline
\end{tabular}
\end{ruledtabular}
\end{table*}

The lattice dynamics is investigated by the density functional perturbation theory (DFPT)\cite{123} within the theory of linear
response. This method includes calculations of charge response to the
lattice distortions (allowed by the symmetry operations) for the specified
vectors in the first Brillouin zone. Calculations start from
the previously calculated ground state atomic and electronic configuration and
continue with the self-consistent calculations of the charge response for each different
displacement.  The normal modes of the optical active phonons (in the $\Gamma$ point) are given in Fig.~\ref{fig1}. Because Fe ions are located in the center of inversion of Pnnm space group, they do not contribute to the Raman scattering process,
i.e. the Raman modes of FeSb$_2$ originate only from the Sb atoms vibrations,
in a manner illustrated in Fig.~\ref{fig1}. In the case of
infrared active modes, both the Fe and Sb atoms contribute to the normal modes, see Fig.\ref{fig1}.

\begin{figure*}
\includegraphics[height=0.4\textwidth]{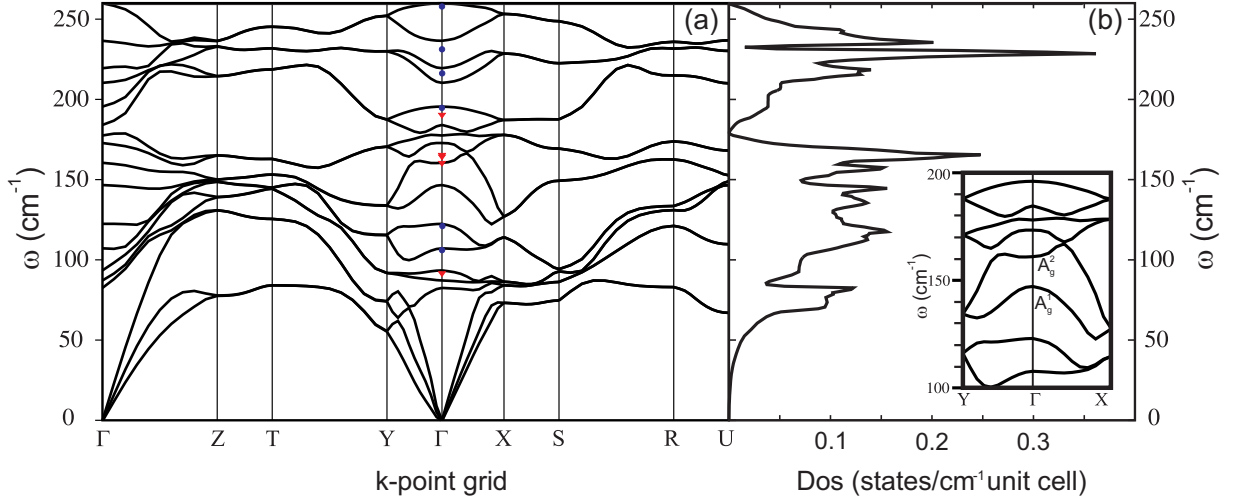}
\caption{Phonon dispersion (a) and phonon density of states (b) for FeSb$_2$. Triangles and circles represent experimentally observed Raman and infrared mode energies. Inset: Dispersion curves of A$_g$ phonon modes along the $\Gamma-Y$ and $\Gamma-X$ directions.}
\label{fig4}
\end{figure*}

In order to obtain the phonon dispersion curves, we have calculated the phonon frequencies at 4$\times$4$\times$4 Monkhorst-Pack $q$-points
mesh and interpolated along the chosen path. Figure \ref{fig4}(a) shows the calculated phonon dispersions, whereas Fig. \ref{fig4}(b) represents the phonon density of states of FeSb$_2$. It is interesting to note that there is a frequency gap in the phonon
dispersion of FeSb$_2$ at about 175 cm$^{-1}$. The lower frequency
range is dominated by Sb-atoms vibrations (mostly Raman active
vibrations), whereas the Fe atoms vibrate at frequencies higher than 175
cm$^{-1}$. These modes are only infrared active.

The phonon density of states peaked structure between 50 an 90 cm$^{-1}$ correspond to the low frequency acoustic modes associated with the low-lying B$_{2g}$ Raman active mode, which calculated frequency is at 89.4 cm$^{-1}$. Sharp peaks in the phonon density of states above 90 cm$^{-1}$ come from the flat regions of dispersion curves of corresponding Raman (below 175 cm$^{-1}$) and infrared (above 175 cm$^{-1}$) modes.

The lattice dynamics calculations allow us to assign the infrared active modes, experimentally observed in Ref.~\onlinecite{10.1}. The assignment of the infrared active modes is done according to the mode energy and symmetry. As we have already mentioned, for $E||b$ (B$_{3u}$ symmetry modes) four modes are observed\cite{10.1} instead of three. We believe that the appearance of two modes at about 257 and 271 cm$^{-1}$ instead of a single frequency mode is the consequence of splitting of relatively broad oscillator (which calculated TO frequency is 259.6
cm$^{-1}$), due to anharmonicity effects.\cite{52} The B$_{1u}$ infrared active mode of FeSb$_2$ is recently observed at
195 cm$^{-1}$ in Ref.~\onlinecite{10.2}.The frequencies and assignment of all infrared active modes are given in Table~\ref{tab.1}.

The DFT calculations are performed at zero
temperature and should be matched with the phonon energies at zero
temperature. For this purpose, we have analyzed the change of the Raman mode energy and linewidth with temperature, induced by anharmonicity effect.

The influence of the anharmonic effects on the Raman mode energy can be taken into account via three- and four-phonon processes by applying the Klemens's ansatz:\cite{17,Klemens}
\begin{equation}
\begin{gathered}
\Omega (T)=\Omega_0-\Delta^{(3)}(T)-\Delta^{(4)}(T),\\
\label{1}
\Delta^{(3)}(T)=C\bigg(1+\frac{2}{e^{x}-1}\bigg),\\
\Delta^{(4)}(T)=D\bigg(1+\frac{3}{e^{y}-1}+\frac{3}{(e^{y}-1)^2}\bigg),
\end{gathered}
\end{equation}
where $\Omega_0$ is the temperature independent contributions to the Raman mode energy, C (D) is the three (four)-phonon
anharmonic constant, $x=\hbar\Omega_0/2k_BT$ and $y=\hbar\Omega_0/3k_BT$.

There are two main contributions to the phonon linewidth: (i) anharmonic decay of the phonon, and (ii) perturbation of the translational symmetry of the crystal by the presence of impurities and defects. Having this in mind, the phonon linewidth can be described with:
\begin{equation}
\begin{gathered}
\Gamma (T)=\Gamma_0+\Gamma^{(3)}(T)+\Gamma^{(4)}(T),\\
\label{2}
\Gamma^{(3)}(T)=A\bigg(1+\frac{2}{e^{x}-1}\bigg)\\
\Gamma^{(4)}(T)=B\bigg(1+\frac{3}{e^{y}-1}+\frac{3}{(e^{y}-1)^2}\bigg)
\end{gathered}
\end{equation}
where $\Gamma_0$ is the temperature independent linewidth, which originates mainly from (ii), A (B) is the three (four)-phonon
anharmonic constant.
\begin{figure}
\includegraphics[height=0.43\textwidth]{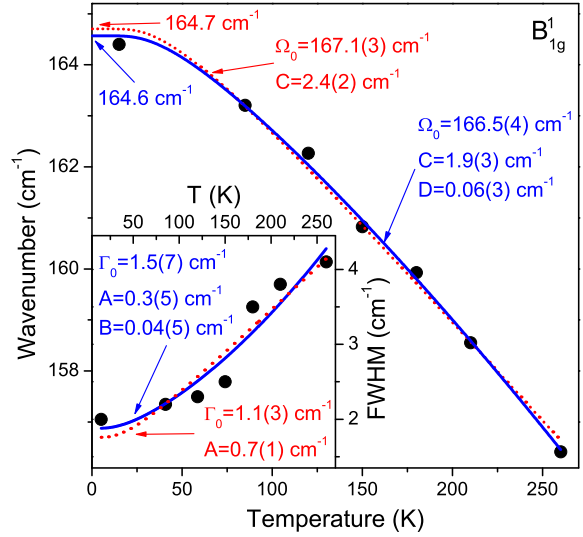}
\caption{(Color online) B$^1_{1g}$ mode wavenumber as a function of temperature.\ Solid line represents theoretical
extrapolation by using Eq.~(\ref{1}), whereas dashed line represent theoretical extrapolation obtained by omitting four-phonon contribution in Eq.~(\ref{1}).\ Inset: Theoretical calculation of FWHM obtained by using Eq.~(\ref{2}) (solid line) and  by omitting four-phonon contribution in Eq.~(\ref{2}) (dashed line).} \label{fig7}
\end{figure}
Analysis of energy and FWHM (full width at half maximum) vs. temperature for the B$^1_{1g}$ mode is presented in Fig.~\ref{fig7}. Because anharmonicity constants ratio B/A and D/C is very small, see Fig.~\ref{fig7}, the contribution of the four-phonon processes is small compared to that of the three-phonon processes. The obtained value of $\Omega(0)=$164.6 cm$^{-1}$ for this mode at zero temperature is in good
agreement with the DFT results. Similar analysis have
been performed for B$^2_{1g}$ symmetry mode
giving the value of $\Omega(0)=$190.4 cm$^{-1}$ at zero temperature,\cite{7.1} which is in rather
good agrement with our calculations.

The calculated energy (89.4 cm$^{-1}$) for the
B$_{2g}$ symmetry mode in the $\Gamma$ point shows excellent
agreement with the room temperature experimental data. This is to be expected since low
energy modes show week anharmonicity effects. Surprisingly large
discrepancy between experimental and calculated phonon energies is
observed for the B$_{3g}$ mode. Since the B$_{2g}$ and B$_{3g}$ modes
have similar normal modes, the chain rotation around the x and y axis, respectively (see Fig.~\ref{fig1}) their frequencies should be very close. This large disagreement is also unexpected since all other calculated phonon
energies show rather good agreement with the
experimentally obtained data. By detailed inspection of our
previously published Raman spectra\cite{7} of pure, Co and Cr doped
FeSb$_2$ samples we did not find any mode in a low frequency region
close to the calculated frequency (95 cm$^{-1}$) for the B$_{3g}$ mode. The
missing B$_{3g}$ mode is most probably of a very low intensity and it was
not possible to extract it from the noise. The mode observed at
151.7 cm$^{-1}$ for (x'y) polarization, which we assigned in Ref. \onlinecite{7} as the
B$_{3g}$ mode, could be the "leakage" of the A$^1_{g}$ mode, which appears at about 150.7 cm$^{-1}$ in the ($x'x'$) polarization.

It is interesting to note that the dispersion curves of two A$_g$ symmetry Raman modes have opposites slopes near the $\Gamma$ point (see the inset of Fig.~\ref{fig4}(b)), which leads to the mode mixing with the ''anticrossing'' effect. A$_g^1$ mode represents stretching vibration of Sb ions, whereas A$_g^2$ mode represents twisting of Sb ions which tent to rotate Sb ions around the z-axis, see Fig~\ref{fig1}. In our previous paper,\cite{7.1} we
showed the existence of the A$_g$ mode mixing in the
case of pure FeSb$_2$ and Cr alloyed samples. Here we
present detailed analysis of the mixing of two A$_g$ modes for pure and
25\% Co alloyed samples.
\begin{figure}
\includegraphics[width=0.33\textwidth]{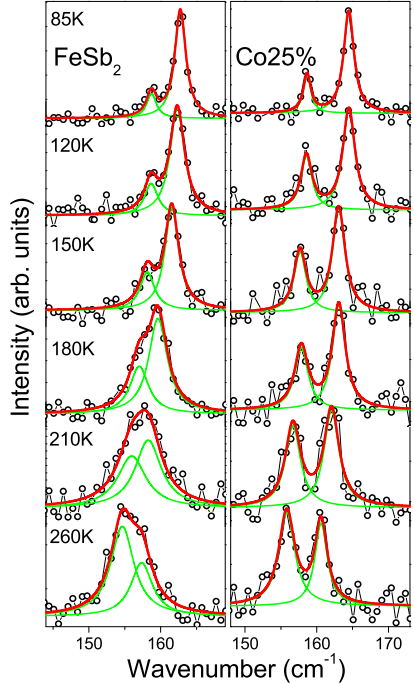}
\caption{(Color online) The Raman scattering spectra of FeSb$_2$ (left
panel)\cite{7.1} and Fe$_{0.75}$Co$_{0.25}$Sb$_2$ (right panel) single crystals in
the (x$^,$x$^,$), x$^,=\frac{1}{\sqrt{2}}[101]$, configuration (A$_{g}$ symmetry modes) measured at various temperatures. }
\label{fig5}
\end{figure}

\begin{figure}
\includegraphics[width=0.45\textwidth]{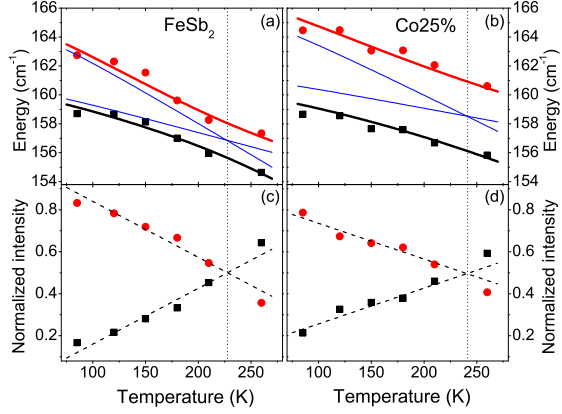}
\caption{(Color online) Energies (a), (b) and normalized intensities
(c), (d) as a function of temperature of the A$_g$ modes for FeSb$_2$
and Fe$_{0.75}$Co$_{0.25}$Sb$_2$ single crystals. Thin solid lines
show energy vs temperature dependence of the A$_g$ modes without
coupling.  Thick solid lines (red and black) show mode energy
temperature dependance for two coupled A$_g$ modes calculated using
Eq. (\ref{3}). The dashed lines are guide to the eye.} \label{fig6}
\end{figure}

The polarized Raman scattering spectra for pure FeSb$_2$ (left panel\cite{7.1})and Fe$_{0.75}$Co$_{0.25}$Sb$_2$ (right panel) single crystals, measured in the $(x'x')$
configuration (A$_g$ modes) at different temperatures, are presented in Fig.~\ref{fig5}. The Lorentzian lineshape profile has been used for the extraction of mode energy and linewidth.
Fig. \ref{fig6} shows the energies and normalized intensities as a
function of temperature of the A$_g$ modes for FeSb$_2$ and
Fe$_{0.75}$Co$_{0.25}$Sb$_2$ single crystals. In the observed
temperature range, energies of two A$_g$ modes for pure and 25\%
Co doped samples are very close which implies the existence of the
mode mixing, manifested by mode repulsion and intensity
transfer with the change of temperature.\cite{13} Indeed, intensities of
these modes are exchanged for both samples in the temperature range
between 210 and 260 K (see Figs. \ref{fig5} and \ref{fig6}).

In general, two phonon branches or any other elementary excitations
of the same symmetry, may couple leading to the renormalization of the
quasiparticle energies. Coupling between two phonon branches yields
to the energy and linewidth changes (anticrossing effect). We
can consider the coupling of two phonon branches as coupling of two
quantum oscillators. When the perturbation is small, we can write the
Hamiltonian of the system as
\begin{equation}
\hat{H}=\left[
  \begin{array}{cc}\vspace{0.2cm}
    \Omega_1(T) & V \\
    V & \Omega_2(T) \\
  \end{array}
\right],
\label{3}
\end{equation}
where $V$ is the interaction constant, $\Omega_1(T)$ and $\Omega_2(T)$ are the unperturbed mode energies, obtained by takeing into account, due to simplicity, only three-phonon process in Eq.~(\ref{1}). The eigenvalues of the Hamiltonian are given by
\begin{equation}
\omega_{\pm}=\frac{1}{2}\left(\Omega_1(T)+\Omega_2(T) \pm \sqrt{(\Omega_1(T)-\Omega_2(T))^2+4V^2} \right).
\label{3}
\end{equation}
Eq.~(\ref{3})  gives a rather good fit of the experimental data (solid lines in Fig.
\ref{fig6} (a),(b)), suggesting the absence of
any additional temperature dependent couplings (i.e. electron-phonon
interaction) for these modes. Fitting parameters are presented in Table \ref{tab.2}.
\begin{table}
\caption{Best fit parameters for energy temperature dependance of the A$_{g}$ symmetry modes using Eq. (\ref{3}).}
\label{tab.2}
\begin{ruledtabular}
\setcellgapes{2pt}
\makegapedcells
\centering
\begin{tabular}{ccccc}
Compound                                          & Symmetry   &$\Omega_0$ (cm$^{-1}$) & $C$ (cm$^{-1}$)   & V (cm$^{-1}$) \\ \hline
\multirow{2}{*}{FeSb$_2$}                         & A$_g^1$    & 167.1                   & 2.65               & \multirow{2}{*}{1.2}     \\
                                                  & A$_g^2$    & 161.5                 & 1.16              &                          \\ \hline
\multirow{2}{*}{Fe$_{0.75}$Co$_{0.25}$Sb$_2$}    & A$_g^1$    & 167.5                 & 2.2               & \multirow{2}{*}{2.4}     \\
                                                  & A$_g^2$    & 161.9                   & 0.80              &

\end{tabular}
\end{ruledtabular}
\end{table}
Zero-temperature energies of A$_g^1$ and A$_g^2$ symmetry modes, in the absence of interaction, for pure (25\% Co dopped) sample are  $\Omega_1(0)=160.3$ cm$^{-1}$ and $\Omega_2(0)=164.5$ cm$^{-1}$ ($\Omega_1(0)=161.1$ cm$^{-1}$ and $\Omega_2(0)=165.3$ cm$^{-1}$). One can notice that the zero-temperature energies for decoupled modes are increased by 0.8 cm$^{-1}$ (about 0.5\% increase) with 25\% Co dopping, corresponding to the unit cell volume contraction.\cite{4} The phonon energy of the bond-stretching mode scales as $R^{-3}$, where $R$ is the bond length.\cite{51} Since the change in $R^{-3}$ is proportional to
the inverse volume change, we can expect the phonon-energy change for bond-stretching modes (A$_g$ modes) to be inversely proportional to the volume change. Because the Co atom substitutes Fe atom, which is located in the center of the inversion, there is no change in Raman spectra due to the mass effect. Additional repulsion between the coupled modes are due to the interaction. With Co doping, the interaction constant $V$ increases, resulting in larger mode separation for the 25\% dopped sample.

\section{Conclusion}

In summary, we presented a detailed theoretical and experimental study of the FeSb$_2$ phonon dynamics. All experimentally observed Raman and infra-red active modes were successfully assigned. The calculated phonon frequencies in the $\Gamma$ point agree with the measured frequencies. We believe that the low energy B$_{3g}$ mode is of a very low intensity and therefore is not observed in the Raman experiments. The phonon mode at 150.7 cm$^{-1}$, which we previously assigned as the B$_{3g}$ mode, could be the "leakage" of the A$_g^1$ mode. The strong intensity exchange of the A$_g$ symmetry modes, observed in  our Raman scattering experiments in the temperature range between 210 K and 260 K, is successfully described by a simple model of coupling of two phonon branches with the same symmetry. The mode mixing is also implied from the calculated dispersion curves, which show opposite slopes for two A$_g$ modes near the $\Gamma$ point. We find that doping of FeSb$_2$ with Co increases the A$_g$ modes repulsion.

\section*{Acknowledgment}

This work was supported by the Serbian Ministry of Education and Science under Projects ON171032, III45018,
ON171017. Part of this work (C. P. and R. H.) was carried out at the Brookhaven National Laboratory which is operated for the Office of Basic Energy Sciences, U.S. Department of Energy by Brookhaven Science Associates (DE-Ac02-98CH10886). Numerical simulations were run on the AEGIS e-Infrastructure, supported in part by FP7 projects EGI-InSPIRE, PRACE-1IP and HP-SEE.
Z.V.P. and M.M.R. acknowledge support from the Swiss National Science
Foundation through the SCOPES Grant No. IZ73Z0-128169.

$^{\ast }$ Present address: Department of Physics, University of Maryland, College Park MD 20742-4111, USA.


\begin{thebibliography}{99}

\bibitem{2} C. Petrovic, J. W. Kim, S. L. Bud'ko, A. I. Goldman P. C.
Canfield, W. Choe, and G. J. Miller, Phys. Rev. B \textbf{67}, 155205 (2003).

\bibitem{5} C. Petrovic, Y. Lee, T. Vogt, N. D. Lazarov, S. L. Bud'ko, and P.
C. Canfield, Phys. Rev. B \textbf{72}, 045103 (2005).

\bibitem{4} Rongwei Hu, V. F. Mitrovi\'{c}, and C. Petrovic, Phys. Rev. B
\textbf{74}, 195130 (2006).

\bibitem{6} Rongwei Hu, V. F. Mitrovi\'{c}, and C. Petrovic, Phys. Rev. B
\textbf{76}, 115105 (2007).

\bibitem{1} A. Bentien, S. Johnsen, G. K. H. Madsen, B. B. Iversen, and F.
Steglich, Europhys. Lett. \textbf{80}, 39901 (2007).

\bibitem{3} Rongwei Hu, V. F. Mitrovi\'{c}, and C. Petrovic, Appl. Phys. Lett.
\textbf{92}, 182108 (2008).

\bibitem{3.1} P. Sun, N. Oeschler, S. Johnsen, B. B. Iversen, and F. Steglich, Dalton Trans. 39, 1012 (2010).

\bibitem{3.2} Peijie Sun, Niels Oeschler, Simon Johnsen, Bo Brummerstedt Iversen, and Frank Steglich, Phys. Rev. B 79, 153308 (2009).

\bibitem{3.3} H. Takahashi, Y. Yasui, I. Terasaki and M. Sato, J. Phys. Soc. Japan 80, 154708 (2011).

\bibitem{6.1} J. M. Tomczak, K. Haule, T. Miyake, A. Georges, and G. Kotliar, Phys. Rev. B
\textbf{82}, 085104 (2010).

\bibitem{10.1} A. Perucchi, L. Degiorgi, Rongwei Hu, C. Petrovic, and V.F. Mitrovi\'{c}, Eur. Phys. J. B \textbf{54}, 175 (2006).

\bibitem{7.a} H. D. Lutz and B. M\"{u}ller, Phys. Chem. Miner. \textbf{18},
265 (1991).

\bibitem{7.b} A. M. Racu, D. Menzel, J. Schoenes, M. Marutzky, S. Johnsen, and
B. B. Iversen, J. Appl. Phys. \textbf{103}, 07C912 (2008).

\bibitem{7} N. Lazarevi\'{c}, Z. V. Popovi\'{c}, Rongwei Hu, and C. Petrovic,
Phys. Rev. B \textbf{80}, 014302 (2009).

\bibitem{7.1} N. Lazarevi\'{c}, Z. V. Popovi\'{c}, Rongwei Hu, and C. Petrovic,
Phys. Rev. B \textbf{81}, 144302 (2010).

\bibitem{31} H. Holseth, and A. Kjekshus, Acta Chem. Scand. \textbf{22}, 3273 (1968).

\bibitem{32} H. Holseth, A. Kjekshus, and A. F. Andresen, Acta Chem. Scand. \textbf{24}, 3309 (1970).

\bibitem{21.1} http://www.quantum-espresso.org/pseudo.php.

\bibitem{21} P. Giannozzi, S. Baroni, N. Bonini, M. Calandra, R. Car, C. Cavazzoni, D. Ceresoli, G. L Chiarotti, M. Cococcioni, I. Dabo, A. Dal Corso, S. de Gironcoli, S. Fabris, G. Fratesi, R. Gebauer, U. Gerstmann, C. Gougoussis, A. Kokalj, M. Lazzeri, L. Martin-Samos, N. Marzari, F. Mauri, R. Mazzarello, S. Paolini, A. Pasquarello, L. Paulatto, C. Sbraccia, S. Scandolo, G. Sclauzero, A. P Seitsonen, A. Smogunov, P. Umari, and R. M Wentzcovitch, J. Phys. Condens. Matter \textbf{21}, 395502 (2009).

\bibitem{21.2} N. Marzari, D. Vanderbilt, A. De Vita, and M. C. Payne, Phys. Rev.
Lett. \textbf{82}, 3296 (1999).


\bibitem{6.a} A. V. Lukoyanov, V. V. Mazurenko, V. I. Anisimov, M. Sigrist, and T. M. Rice,
Eur. Phys. J. B 53, 205 (2006).

\bibitem{6.b} A. Bentien, G. K. H. Madson, S. Johnsen, and B. B.
Iversen, Phys. Rev. B 74, 205105 (2006).

\bibitem{123} S. Baroni, S. de Gironcoli, A. Dal Corso, and P. Giannozzi, Rev. Mod. Phys. \textbf{73}, 515
(2001).

\bibitem{52} F. Gervais and B. Piriou, Phys. Rev. B \textbf{10}, 1642 (1974).

\bibitem{10.2} A. Herzog, M. Marutzky, J. Sichelschmidt, F. Steglich, S. Kimura, S. Johnsen, and B. B. Iversen,
Phys. Rev. B \textbf{82}, 245205 (2010).

\bibitem{17} M. Balkanski, R. F. Wallis, and E. Haro, Phys. Rev. B \textbf{28}, 1928 (1983).

\bibitem{Klemens}  P. G. Klemens, Phys. Rev. \textbf{148}, 845 (1966).

\bibitem{13} M. N. Iliev, M. V. Abrashev, J. Laverdi�re, S. Jandl, M. M. Gospodinov, Y.-Q. Wang, and Y.-Y. Sun, Phys. Rev. B \textbf{73}, 064302 (2006).

\bibitem{51} Z. V. Popovi\'c, V Stergiou, Y. S. Raptis, M. J. Konstantinovi\'c, M. Isobe, Y. Ueda, and V. V. Moshchalkov, J. Phys.: Condens. Matter \textbf{14}, L583 (2002).





\end{thebibliography}
\end{document}